# EVIDENCE FOR GLUON RECOMBINATION IN DEEP INELASTIC SCATTERING

Kjell Prytz
University of Gavle
801 76 Gavle
Sweden
e-mail: kjell.prytz@hig.se

**Abstract**
The pomeron structure function is extracted from the latest H1 data and are subject to a QCD analysis. The result shows evidence for gluon recombination.

**Introduction**
We discuss the interpretation of the diffractive structure function as measured by the H1 collaboration at DESY [1]. Using a model proposed in ref. 2 we extract the pomeron structure function, a notion that occur in Regge phenomenlogy [3]. Together with the Collins factorization theorem for diffractive scattering [4] a framework for applying QCD evolution equations to the pomeron parton dynamics can be constructed.
The procedure involves many assumptions but since it was first proposed some 10-15 years ago [2] evidence and support for its correctness have been presented (see ref . 5 and references therein).
In 1997, H1 published a paper [6] where the DGLAP QCD dynamics [7] was applied to their own first data. Their result showed that the DGLAP evolution equations are able to fit these data satisfactorily. However, in 1998 new data with an extended kinematic range were published [8] and it was found that the DGLAP equations could not provide a decent description of the data [1].
In ref. 5 it was explained why DGLAP doesn't work for the latest H1 data and it was *qualitatively* shown that the inclusion of gluon recombination (or generally, a low-x higher twist term) in the analysis is sufficient to explain the data. These equations are denoted GLR-MQ [9] and corrects the DGLAP equations by adding terms accounting for gluon recombination.
In this paper we apply the GLR-MQ equations *quantitatively* to the data by performing a full QCD analysis of the H1 data in leading order. The sole aim is to check whether the DGLAP equations can be ruled out in favour of the GLR-MQ equations.

**Pomeron Phenomenology**
In our model, the electromagnetic diffractive process is assumed to occur as depicted in fig. 1.

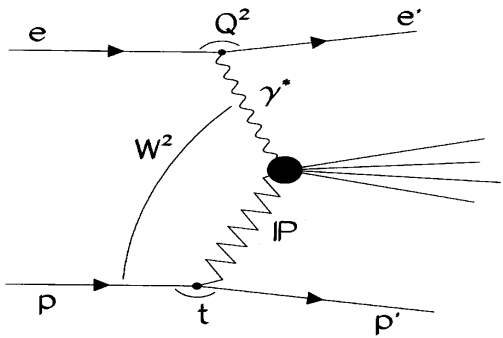

Figure 1: The diffractive process with pomeron exchange in electromagnetic electron-proton high energy scattering.

There are three vertices which are subject to the factorization theorems of Collins (upper) and of Regge (lower). The electromagnetic cross section can approximately be written [2]

$$\frac{d\sigma(ep \to epX)}{dx_{Pom}\,dt\,dx\,dQ^2} = \frac{4\pi\alpha^2}{xQ^2}\left[1 - y + \frac{y^2}{2}\right] F_2^D(x_{Pom}, t, x, Q^2) \qquad (1)$$

where $F_2^D$ is called the diffractive structure function and the corresponding $F_1^D$ structure function has been neglected.

In the laboratory frame, $y = \nu/E$ where $\nu$ is the energy transferred by the photon in the process and E is the energy carried by the incoming electron. $Q^2 = -(k-k')^2$ where k and k' are the momenta of the incoming and outgoing electron respectively and $x = Q^2/2M\nu$ where M is the proton mass. $\alpha$ is the electromagnetic coupling, t is the squared four-momentum lost by the proton and $x_{Pom}$ is the fraction of the incoming proton momentum lost by the proton which in the model is the fractional momentum carried by the pomeron.

The Regge factorization theorem then leads to [2]

$$F_2^D(x_{Pom}, t, x, Q^2) = f_{Pom/p}(x_{Pom}, t)\, F_2^{Pom}\!\left(\frac{x}{x_{Pom}}, Q^2\right) \qquad (2)$$

were $f_{Pom/p}$ is usually called "the pomeron flux" and describes the coupling between the pomeron and proton. In a "quark-parton model jargong" one would say that $f_{Pom/p}$ measures a (unnormalized) probability to find a pomeron "inside" the proton [2].

We introduce $\beta = x/x_{Pom}$ which is interpreted (in the infinite momentum frame) as the fraction of the pomeron momentum carried by the struck parton.

H1 presents their structure function data in terms of $x_{Pom} F_2^D$. In order to obtain the pomeron structure function ($F_2^{Pom}$), which will be subject to the QCD analyses, we need to divide by $x_{Pom} f_{Pom/p}$.

For the pomeron flux we use the information obtained by H1 [6]:

$$f_{Pom/p}(x_{Pom}, t) = \frac{e^{bt}}{x_{Pom}^{2\alpha_{Pom}-1}} \qquad (3)$$

where $b = 4.6$ GeV$^2$ and $\alpha_{Pom}(t) = \alpha_{Pom}(0) + \alpha'_{Pom}\, t = 1.2 + 0.26\,t$ is the Regge trajectory for the pomeron. An integration over t is performed from $-1.0$ GeV$^2$ to 0 and $x_{Pom} = 0.005$. At this low value of $x_{Pom}$, H1 showed that there is negligible contribution from other exchange particles beyond the pomeron [6].

We use the following evolution equations

$$\frac{dG(\beta, Q^2)}{d\log Q^2} = \frac{\alpha_s(Q^2)}{2\pi}\,\beta\left[\int_\beta^1 \frac{dy}{y^2} G(y, Q^2) P_{gg}\!\left(\frac{\beta}{y}\right) + \int_\beta^1 \frac{dy}{y^2} y\, q(y, Q^2) P_{gq}\!\left(\frac{\beta}{y}\right)\right]$$

$$- \frac{81\alpha_s^2(Q^2)}{16 R^2 Q^2}\int_\beta^1 \frac{dy}{y}[G(y,Q)]^2 + \frac{81\alpha_s^2(Q^2)}{16 R^2 Q^2}\int_{\beta/2}^\beta \frac{dy}{y}[G(y,Q)]^2 \qquad (4)$$

$$\frac{d\beta q(\beta,Q^2)}{d\log Q^2} = \frac{\alpha_s(Q^2)}{2\pi} \beta \left[ \int_\beta^1 \frac{dy}{y^2} G(y,Q^2) P_{qg}\left(\frac{\beta}{y}\right) + \int_\beta^1 \frac{dy}{y^2} yq(y,Q^2) P_{qq}\left(\frac{\beta}{y}\right) \right]$$
$$- \frac{27\alpha_s^2(Q^2)}{160 R^2 Q^2}[G(\beta,Q^2)]^2 + HT \quad (5)$$

where on the right hand side, the terms linear in gluon distribution G and quark distribution yq are given by DGLAP [7] and the terms quadratic in G are given by GLR-MQ [9]. The latter terms account for gluon recombination into gluons (eq. 1) and into quarks (eq. 2).
The fourth term in the gluon evolution equation was introduced in order to conserve gluon momentum [10]. It was achieved by simply noting that a fusion of a couple of gluon results in a new one with higher momentum.
The amount of gluon recombination is controlled by the pomeron size parameter R. The term denoted 'HT' was introduced by Mueller and Qiu [9] but not fully given. It was neglected in our calculation.
For the initial gluon and quark distributions, given at the lowest used $Q^2 = 1.2\ \text{GeV}^2$, we use the following forms:

$$\beta q = C_1 + C_2 \beta + C_3 \beta^2 + C_4 \beta^3 + C_5 \beta^4 + C_6 \beta^5$$
$$G = \beta g = C_7 \beta^{C_8} (1-\beta)^{C_9}$$

The rather complicated form of the quark distribution is dictated by data.
The pomeron is assumed to be a flavour singlet particle so that the evolution is purely singlet. We use four flavours and the QCD parameter $\Lambda = 200$ MeV.
We adopt a so called saturation procedure [11] meaning that if, in the gluon evolution, the sum of the recombination terms get larger than the DGLAP term we set the gluon derivative to zero. This actually happens in the low-$\beta$, low-$Q^2$ region of the data (see fig. 3) so a more thorough theoretical investigation about this effect is asked for.

## Results

The result of the fit using the GLR-MQ equations is shown in fig. 2 with $\chi^2$/dof = 58/62.

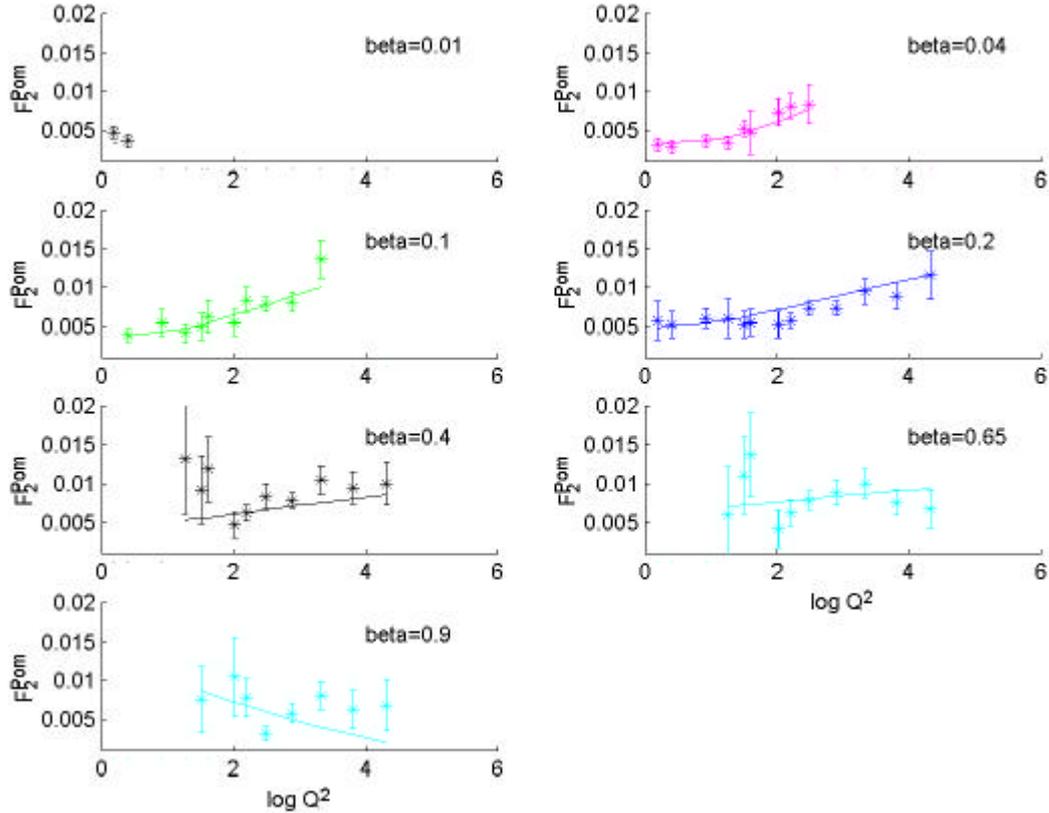

Figure 2: *Optimized fit of the H1 data based on the GLR-MQ equations.*

There are three free quantities in the fit: the quark and gluon distributions and the pomeron size. They mainly control the level, the $Q^2$ dependence and the flattening of the data at low-$Q^2$ respectively. There is no size parameter in the DGLAP equations and accordingly they cannot account for the "flattening effect" seen at low-$Q^2$.
The result on the quark and gluon distribution is shown in fig. 3.

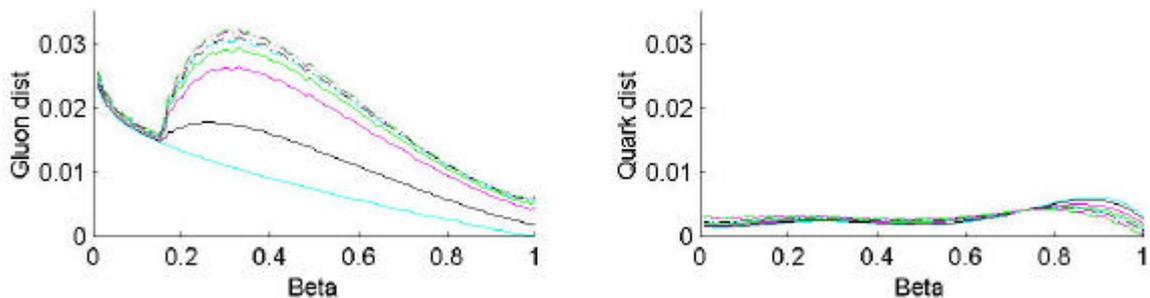

Figure 3: *The resulting gluon and quark distribution from the fit based on the GLR-MQ equations. The lines corresponds to $Q^2$ =1.2, 1.5, 2.5, 3.5, 4.5, 5, 7.5 and 9 GeV$^2$.*

As has been noted several times earlier the gluons dominate in the pomeron. As can be seen in fig. 3, the gluon distribution exhibits an unusual evolution at low-$\beta$. This is due to the saturation procedure adopted (see above). We expect that terms of twist-6 and higher, still missing in the theory, would make the distribution smoother.

Furthermore, we note that the momentum sum rule is not fulfilled and momentum is not even conserved. The latter is mainly due to the saturation effect and the former has been discussed several times previously in the literature. In our view this is a theoretical issue which needs more considerations.

As for the pomeron size we obtain R = 0.03 GeV$^{-1}$ which is considerably smaller than previously obtained [12]. But since we don't aim for any error and correlation analysis of the fitting parameters in this paper we wish to refrain from drawing any conclusions from this. It should at this moment only be considered as a fit parameter. Clearly, this parameter is highly correlated with the gluon density and probably we need an extended $Q^2$ lever arm in order to disentangle the correlation satisfactorily.

We also perform a fit based on the DGLAP scheme shown in fig. 4 and resulting in $\chi^2$/dof = 72/62. In view of the rather large errors of the data, we consider this as a poor fit.

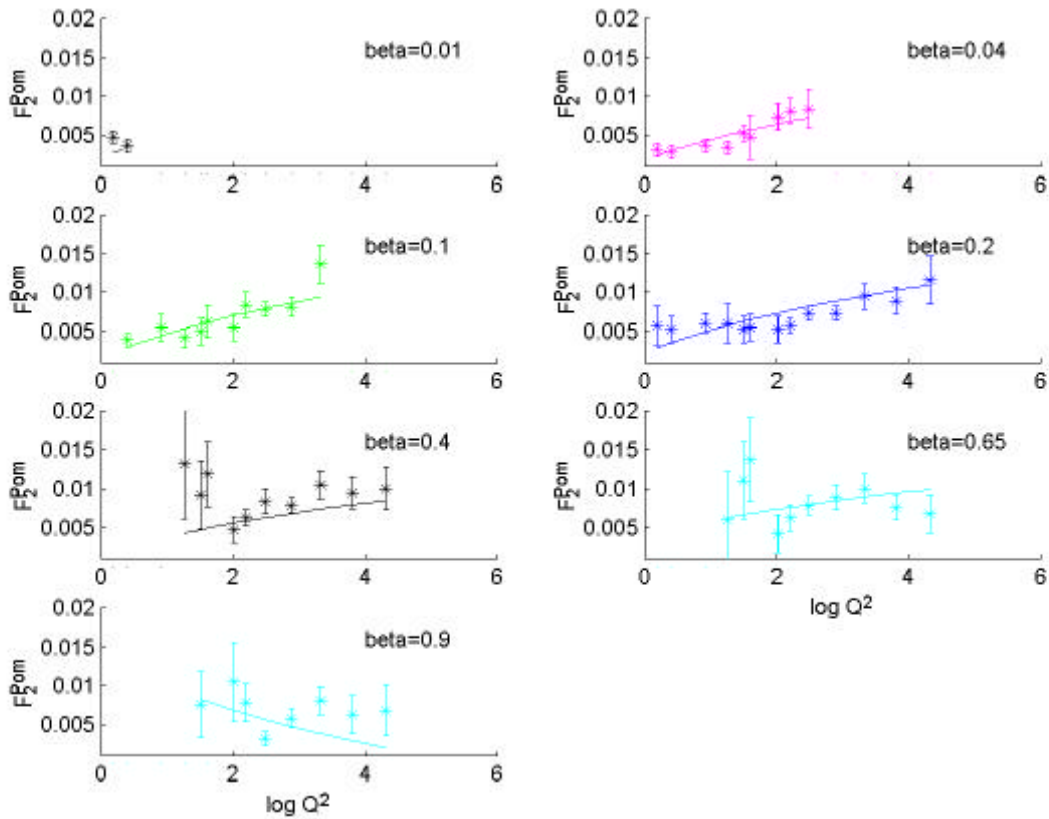

Figure 4: *Optimized fit of the H1 data based on the DGLAP equations.*

The resulting parton distributions are shown in fig. 5.

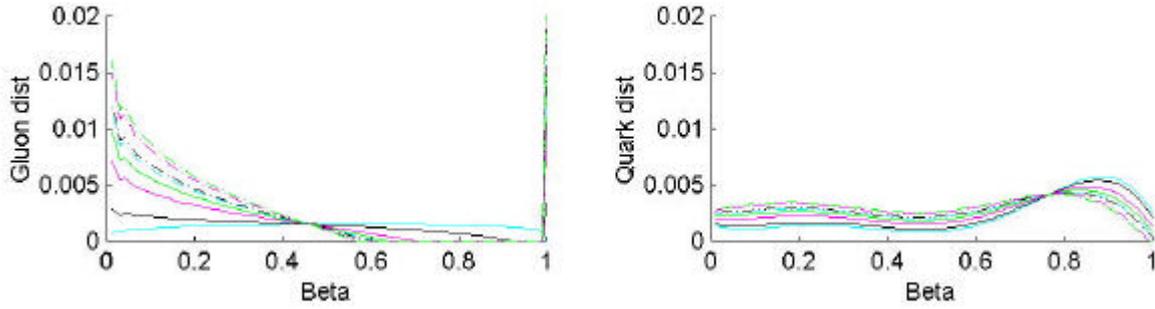

Figure 5: *The resulting gluon and quark distribution from the fit based on the DGLAP equations. The lines corresponds to $Q^2 = 1.2, 1.5, 2.5, 3.5, 4.5, 5, 7.5$ and $9\ GeV^2$.*

Since we look for gluon recombination in this analysis the region of low $\beta$ is most interesting. We therefore in addition investigate the fit quality for the first four $\beta$-bins only, where also the data quality is best. For the GLR-MQ dynamics (figure 2) we then obtain $\chi^2/dof = 20/34$ whereas for the DGLAP dynamics (figure 4) we get $\chi^2/dof = 33/34$. This indicate a favour for the GLR-MQ equations. Note that the errors on the data include systematics so that $\chi^2/dof < 1$ is expected for a perfect fit.

**Discussion**
We have seen that a large contribution of gluon recombination is needed in order to describe the data. Being used to discuss this effect in terms of 'low-x effects' it might be surprising to see that the whole range of $\beta$ in the gluon evolution (compare fig. 3 and 5) is affected. However, the reason for associating gluon recombination with 'low-x physics' is because it was first discussed in connection with nucleon structure. What really governs this dynamic is the gluon spatial density which is high at low-x in the nucleon. In the pomeron, however, due to its small size and gluon domination its gluon density is high throughout the kinematical region and gluon recombination is a natural part of its whole QCD evolution.
One can of course question the whole approach of applying the GLR-MQ equations under such circumstances since we would expect even higher twist terms to become influential. These terms are not yet available but at the moment they serve as an explanation for the fact that the gluon distribution obtained in fig. 3 is rather odd, due to the kink at $\beta=0.15$, and also that the pomeron size obtained is much smaller than expected. Nevertheless, although we could expect a more acceptable result when including terms of this kind, we certainly claim that the main interpretation of the data is given in this analysis; the pomeron is gluon dominated, exhibiting gluon recombination at such a level that saturation occurs in the kinematical region of the data and that in this model the pomeron is much smaller than the nucleon.
After all, we cannot deny the data where the huge effect of gluon recombination is visible already 'by eye'. Also, the effect of saturation is indicated through the flattening of data at low-$Q^2$.
We can also discuss and question the numerical result on the pomeron size which is, as mentioned above, strongly correlated with the gluon distribution. Due to this correlation we cannot increase the pomeron size since that just results in an increasing gluon density in order to keep the amount of gluon recombination approximately unchanged. As a consequence, the DGLAP terms will then increase contributions to the positive scaling violations and ruin the fit. Alternatively, one could try to account for the flattening effect with the DGLAP dynamics only and expect a very low gluon density at low-$\beta$. This was done by H1 [6] with their old data but it doesn't work with the present data set since the optimum gluon distribution is shown in figure 5. Indeed, the density is low but not low enough to get a good fit.

We should mention though, that we haven't made any attempt to allow for negative values of the gluon distribution. Such an approach could very well change the conclusions since it may account for the flattening effect at low-$Q^2$. However, our opinion is that this is unphysical since our basis has been the physical quark-parton model. Since this model can account for the data when including gluon recombination we see no reason for such an exotic adventure.

**Conclusions and outlooks**

We have for the first time found a model which fit the complete H1 diffractive data set for $x_{Pom} = 0.005$. We have shown that the inclusion of gluon recombination solves the problem if we at the same time introduce an effect of gluon saturation.

The parton densities found accordingly confirm the gluonic domination in the pomeron but the shape of the gluon distribution differs from that found by H1 [6] (fit 3).

Our investigation also points to the needs of measurements extending the kinematical range of the data. We could then learn more about gluon recombination and in particular get safer information on the pomeron size.

The recent preliminary data from H1 [13] might add further information concerning gluon recombination mainly due to improved precision. The general trend, however, is identical with the data used here, i.e. for $F_2$ vs $Q^2$ we see a flattening effect at low $Q^2$ for the lowest $\beta$ bins at $x_{Pom} < 0.005$. Data at higher $x_{Pom}$ is not considered by us due to contribution of other exchange particles than the pomeron.

In the H1 paper [13] they claim that the DGLAP equations can describe the data. I disagree with this conclusion. No $\chi^2$ is given but one can see bye eye that the flattening effect is not taken care of (see the bins $\beta=0.1$ and $\beta=0.2$ in figure 9 of this paper). No attempt to fit with GLR-MQ is made. In this analysis minimum $Q^2 = 6$ GeV$^2$ whereas the most pronounced effects of gluon recombination occur at lower $Q^2$ explaining why they obtain an apparent reasonable DGLAP fit.

As soon as these data are presented in tables one should use them to perform the same fits as in this paper. Since these data seem to agree with those that have been used in this paper, differing just in precision, and since no kinematical extension is presented, we cannot expect any other kind of conclusion than what has been drawn here.

**Acknowledgements**

Many thanks to Gunnar Ingelman for helpful suggestions. The work has been supported economically by the University of Gavle, Sweden.